\documentclass[conference]{IEEEtran}
\IEEEoverridecommandlockouts
\usepackage{cite}
\usepackage{amsmath,amssymb,amsfonts}
\usepackage{algorithmic}
\usepackage{graphicx}
\usepackage{textcomp}
\usepackage{xcolor}
\usepackage{url}
\def\BibTeX{{\rm B\kern-.05em{\sc i\kern-.025em b}\kern-.08em
    T\kern-.1667em\lower.7ex\hbox{E}\kern-.125emX}}
\begin{document}

\title{Analysis of Traffic Congestion in North Campus, Delhi University Using Continuous Time Models}

\author{\IEEEauthorblockN{Siddhartha Mahajan}
\IEEEauthorblockA{\textit{Cluster Innovation Centre} \\
\textit{University Of Delhi}\\
siddharthamahajan03@gmail.com}
\and

\IEEEauthorblockN{Harsh Raj}
\IEEEauthorblockA{\textit{Cluster Innovation Centre} \\
\textit{University Of Delhi}\\
rajharsh1810@gmail.com}
\and

\IEEEauthorblockN{Dr. Sonam Tanwar}
\IEEEauthorblockA{\textit{Cluster Innovation Centre} \\
\textit{University Of Delhi}\\
sonaiitr@gmail.com}
}

\maketitle

\begin{abstract}
This project investigates traffic congestion within North Campus, Delhi University (DU), using continuous time simulations implemented in UXSim to model vehicle movement and interaction. The study focuses on several key intersections, identifies recurring congestion points, and evaluates the effectiveness of conventional traffic management measures. Implementing signal timing optimization and modest intersection reconfiguration resulted in measurable improvements in simulated traffic flow. The results provide practical insights for local traffic management and illustrate the value of continuous time simulation methods for informing short-term interventions and longer-term planning.
\end{abstract}

\section{Introduction}
Traffic congestion is a persistent problem in densely used urban precincts and it is especially visible in academic campus environments. North Campus, Delhi University, combines a complex mix of motor vehicles, two-wheelers, pedestrians and periodic surges associated with class dismissals and campus events. These conditions increase travel times, raise local emissions and impose economic costs on commuters and the surrounding community.

This paper presents an analysis of congestion in North Campus using continuous time simulation. Continuous time approaches update vehicle states in response to events rather than at fixed intervals; this allows the model to capture transient dynamics such as queue formation and shock-wave propagation more accurately than many discrete time methods. We implemented the model in UXSim and calibrated it with field survey data. The primary objectives are to produce a detailed network representation of North Campus, to identify recurrent congestion locations and their causes, to evaluate a set of realistic traffic scenarios including class dismissals and events, and to assess conventional traffic management measures such as signal timing adjustments and intersection reconfiguration. The intention is to provide actionable insights that can support traffic management decisions at the campus scale.

\section{Methodology}
This section describes data collection, map preparation, preprocessing steps, the continuous time simulation approach and how UXSim was used to construct and evaluate scenarios.

\subsection{Data collection}
We collected traffic counts, pedestrian flows and contextual observations at a set of critical locations across North Campus. Survey teams followed a predefined protocol to obtain vehicle and pedestrian counts during peak and off-peak periods as well as during special-event periods. Observers used tally counters, stopwatches and GPS-enabled devices for time stamping and positional verification. Manual counts were cross-checked against digital recordings to identify and correct discrepancies. Surveys were repeated across multiple days to capture variability, and simultaneous teams were used at adjacent sites to ensure internal consistency and to reduce measurement error.

\subsection{Map selection and building}
\begin{figure}[h]
    \centering
    \includegraphics[width=0.45\textwidth]{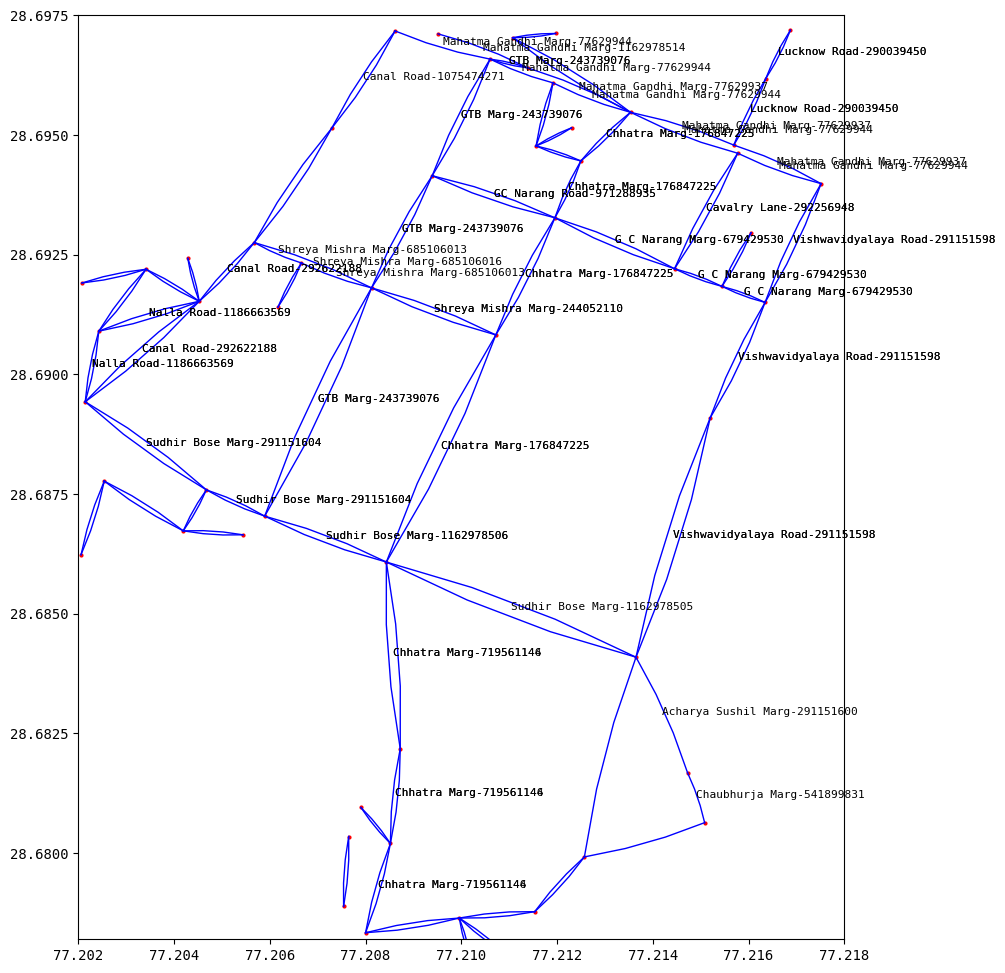}
    \caption{Final cleaned map of the North Campus road network used in the simulation.}
    \label{fig:map}
\end{figure}

A geographically accurate network is essential for realistic simulation. We extracted road geometry and topology from OpenStreetMap and confined the study area to the coordinates (77.202, 28.6782) through (77.218, 28.6975), which encompass the primary campus roads and intersections. The raw map data were cleaned to remove redundant nodes and to correct geometric inconsistencies such as overlapping segments and misaligned junctions. Road attributes including speed limits, functional classes and connectivity were standardized to create a consistent representation for the simulation platform. The final cleaned map, including street names and topology used in the model, is shown in Figure~\ref{fig:map}.

\subsection{Data preprocessing}
Prior to simulation, field data were processed to produce consistent demand inputs. Counts from multiple sessions were aggregated to estimate typical flow rates at key approaches. Temporal variations were normalized so that different intersection inputs could be compared and combined in a coherent demand model. Automated quality-control routines were applied to flag implausible values and outliers; these cases were reviewed against field notes and corrected where appropriate. Finally, all inputs were reformatted to match UXSim's required data schema so that origin, destination and temporal profiles could be ingested directly by the simulator.

\subsection{Continuous time simulation}
The simulation framework adopts an event-driven continuous time formulation where vehicle positions, speeds and interactions are updated when relevant events occur, such as vehicle arrivals, lane changes or signal phase transitions. This event-based update mechanism provides fine temporal resolution for capturing transient behaviours that often determine congestion onset and dissipation. The model includes representations of lane capacity, turning pockets and pedestrian crossing interactions, allowing detailed analysis of queue dynamics and throughput under varying demand patterns.

\subsection{Application using UXSim}
UXSim was used to instantiate the network, to specify demand and to execute continuous time simulations. Nodes and links were defined using the cleaned map and annotated with geometric and operational attributes such as link lengths, free flow speeds and lane counts. Demand inputs derived from survey data were used to generate origin-destination flows and time-varying arrival profiles. Simulations were run to capture both transient surges and longer steady-state intervals. The platform's visualization tools aided in identifying localized queuing patterns, while exported logs were used to compute performance indicators such as link delays and intersection queue lengths. To evaluate management measures we created alternative network configurations that reflected adjusted signal timings and modest geometric modifications.

\subsection{Traffic scenarios}
We modelled a set of scenarios chosen to represent typical congestion drivers observed on site. Scenario 1 examines the surge that follows classes ending at Ramjas College and covers the immediate post-class hour through the period of gradual stabilization. Scenario 2 focuses on the comparable surge after classes end at Miranda House and emphasizes interactions between dense pedestrian flows and vehicular traffic. Scenario 3 models commuter movements between Malkaganj and Vishvavidyalaya Metro Station and highlights critical links such as GTB Road Bridge and the Mall Road intersection. Scenario 4 represents high-demand periods during college festivals, where visitor influx increases both vehicular and pedestrian loads. Scenario 5 captures examination day patterns, where coordinated surges occur in the morning and afternoon. Each scenario was simulated independently and in sequential combinations to observe cumulative effects and residual congestion propagation.

\section{Results and Discussion}
This section synthesizes the simulation outputs and discusses implications for traffic management.

\subsection{Key congestion points}
The simulations consistently identified a small set of locations as recurrent bottlenecks. The Ramjas-St. Stephen's intersection experienced notable delays during post-class surges, with queue formation on several approaches. The GTB Road Bridge frequently showed high vehicular density that led to sustained slowdowns and queue spillback into upstream links. The Mall Road intersection, where multiple traffic streams converge, exhibited prolonged congestion during peak periods. Areas near the Patel Chest Institute were also vulnerable to buildup during the morning and evening peaks. These hotspots align with field observations and suggest that interventions targeted at these nodes are likely to yield the largest operational benefits.

\subsection{Detailed simulation outcomes}
When classes at Ramjas and Miranda House ended simultaneously, the combined surge produced a sharp increase in queue lengths at adjacent junctions, particularly at the Ramjas-St. Stephen's location. Sequential execution of scenarios revealed that residual traffic from an earlier surge can amplify congestion in later periods, indicating a strong temporal coupling between events. The commuter flow between Malkaganj and Vishvavidyalaya Station repeatedly highlighted GTB Road Bridge and Mall Road as capacity-constrained links, indicating that localized geometric or signal interventions at these points could reduce network-wide delay.

\begin{figure}[h]
    \centering
    \includegraphics[width=0.45\textwidth]{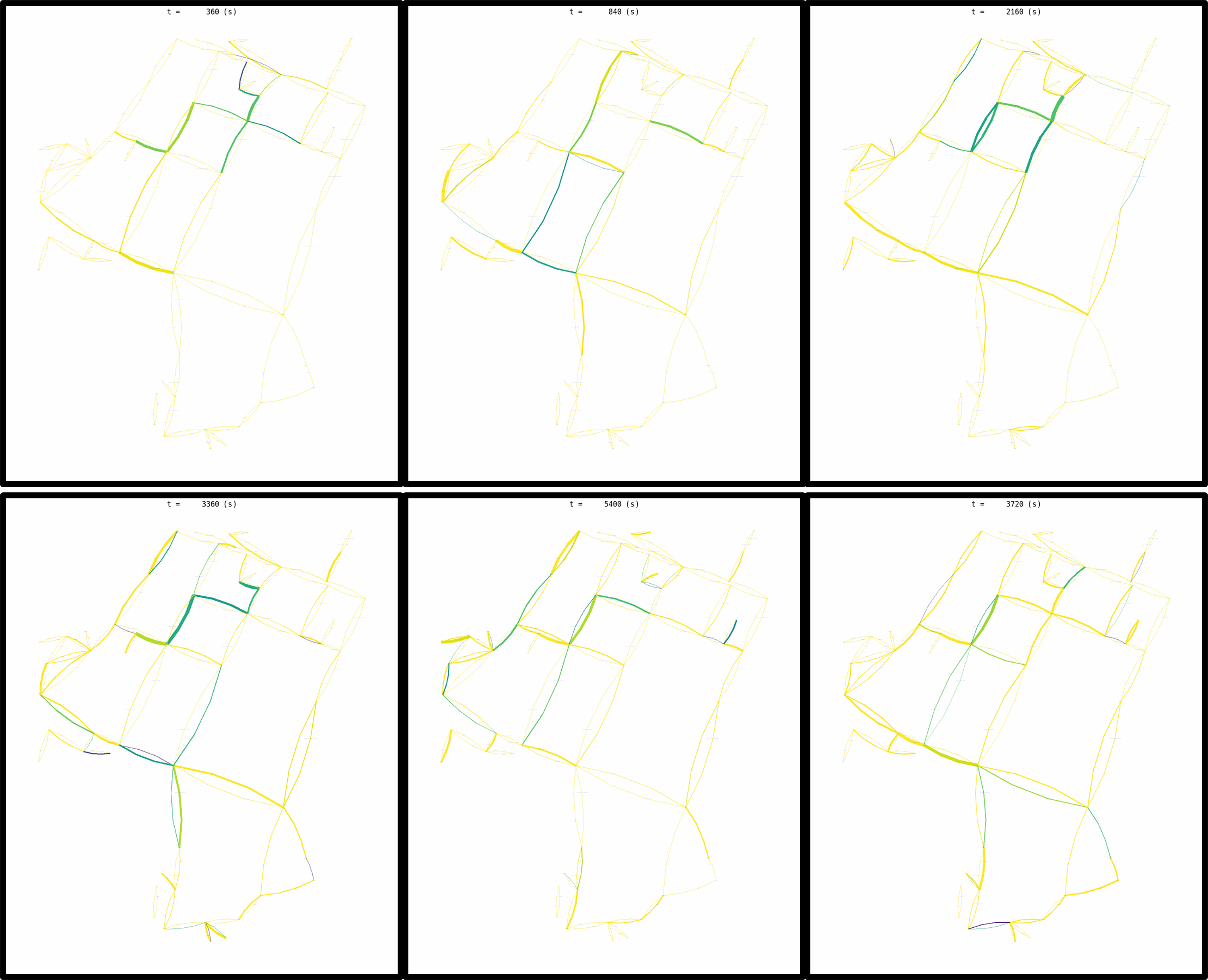}
    \caption{Simulation snapshot of the combined traffic surge at Ramjas and Miranda House.}
    \label{fig:ram_mir}
\end{figure}

\begin{figure}[h]
    \centering
    \includegraphics[width=0.45\textwidth]{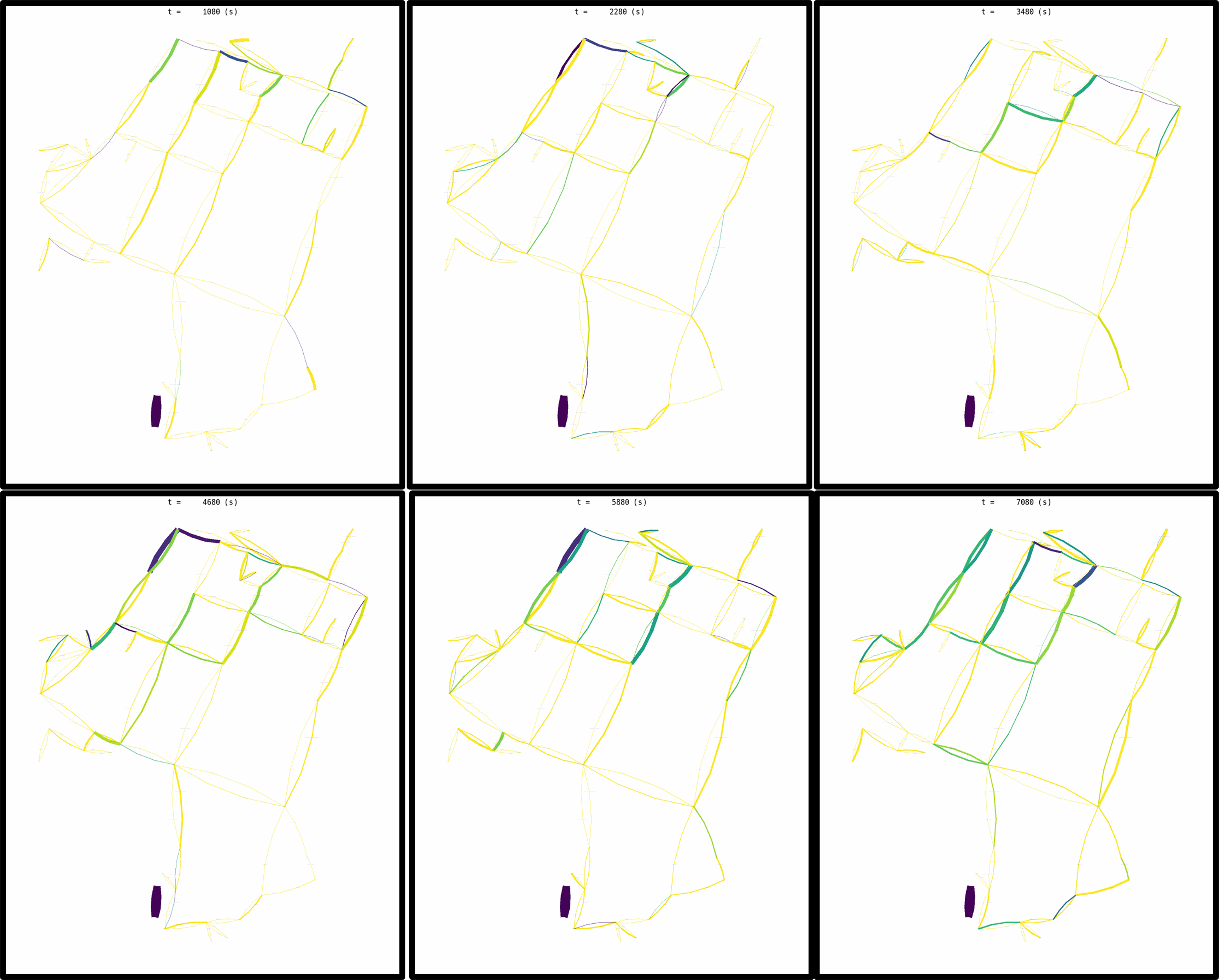}
    \caption{Simulation snapshot for the Malkaganj to Vishvavidyalaya Station corridor.}
    \label{fig:mlk_vv}
\end{figure}

Figure~\ref{fig:ram_mir} shows a representative simulation snapshot of the combined surge at Ramjas and Miranda House, and Figure~\ref{fig:mlk_vv} shows conditions on the corridor between Malkaganj and Vishvavidyalaya Station.

\subsection{Effectiveness of traffic management strategies}
We evaluated conventional measures that are practical to implement at the campus scale. Adjusting signal timing to better match demand profiles consistently reduced waiting times and increased intersection throughput in the simulated runs. Small-scale intersection reconfiguration, such as formalizing turning pockets or refining pedestrian crossing placements, smoothed conflicting movements and reduced queue interference. During event and examination scenarios, implementing coordinated timing offsets and temporary traffic control measures reduced peak accumulation. These results indicate that conventional operational measures, when informed by detailed simulation, can provide meaningful improvements without major infrastructure investment.

\subsection{Limitations and future work}
The study has several limitations that should guide interpretation. Field data were collected using manual and semi-manual methods; integrating continuous sensor feeds would improve the temporal fidelity of demand inputs. The model scope is limited to the North Campus area; extending the geographic boundary would capture additional through movements that can influence local congestion but would require greater computational resources and further calibration. Finally, while this study focused on conventional management measures, future work could explore adaptive control strategies and multi-modal interactions more deeply, including public transport schedules and shared mobility modes.

\section{Conclusion}
This study used continuous time simulation in UXSim, calibrated with field observations, to examine congestion dynamics in North Campus, Delhi University. The analysis identified critical hotspots including the Ramjas-St. Stephen's intersection, GTB Road Bridge and the Mall Road intersection, and demonstrated that operational measures such as signal timing adjustments and modest intersection reconfiguration can reduce delays and improve flow. The results underscore the value of detailed simulation for campus-scale traffic management and provide a baseline for future studies that incorporate real-time data and adaptive control strategies.

\section*{Acknowledgment}
The authors thank the survey teams and campus authorities who facilitated field data collection and provided local contextual knowledge.

\end{document}